\documentclass[runningheads]{llncs}
\usepackage[T1]{fontenc}
\usepackage{graphicx}
\usepackage[numbers]{natbib}  
\usepackage{url} 
\usepackage{lipsum}

\usepackage{caption}
\usepackage{subcaption}
\usepackage{array}
\usepackage{multirow}

\begin{document}
\title{Training Attention Skills in Individuals with Neurodevelopmental Disorders using
Virtual Reality and Eye-tracking technology}
\titlerunning{Training Attention in Individuals with NDD using VR and Eye-tracking}
%
\author{Alberto Patti\inst{1}\orcidID{0000-0002-6871-7417} \and
Francesco Vona\inst{1}\orcidID{0000-0003-4558-4989} \and
Anna Barberio\inst{1*}\orcidID{0009-0005-8451-3923} \and 
Marco Domenico Buttiglione\inst{1*}\orcidID{0009-0000-9202-7486} \and 
Ivan Crusco\inst{1*}\orcidID{0009-0004-3825-5794} \and 
Marco Mores\inst{2}\orcidID{0000-0002-3046-1145} \and 
Franca Garzotto\inst{1}\orcidID{0000-0003-4905-7166}}
\authorrunning{Patti et al.}
%
\institute{Department of Electronics, Information, and Bioengineering, Politecnico di Milano, Milan, Italy \email{name.surname@polimi.it}, \inst{*}\email{name.surname@mail.polimi.it} \url{https://i3lab.polimi.it/}   \and
Fraternità e Amicizia Soc. Coop. Soc. Onlus, Milan, Italy  \email{name.surname@fraternitaeamicizia.it} \url{https://www.fraternitaeamicizia.it/}
}
\maketitle              
\begin{abstract}
Neurodevelopmental disorders (NDD), encompassing conditions like Intellectual Disability, Attention Deficit Hyperactivity Disorder, and Autism Spectrum Disorder, present challenges across various cognitive capacities. Attention deficits are often common in individuals with NDD due to the sensory system dysfunction that characterizes these disorders. Consequently, limited attention capability can affect the overall quality of life and the ability to transfer knowledge from one circumstance to another. The literature has increasingly recognized the potential benefits of virtual reality (VR) in supporting NDD learning and rehabilitation due to its interactive and engaging nature, which is critical for consistent practice. In previous studies, we explored the usage of a VR application called Wildcard to enhance attention skills in persons with NDD. The application has been redesigned in this study, exploiting eye-tracking technology to enable novel and more fine-grade interactions. A four-week experiment with 38 NDD participants was conducted to evaluate its usability and effectiveness in improving Visual Attention Skills. Results show the usability and effectiveness of Wildcard in enhancing attention skills, advocating for continued exploration of VR and eye-tracking technology's potential in NDD interventions.

\end{abstract}
\section{Introduction} 
Neurodevelopmental disorders (NDD) are a group of conditions that typically manifest early in development, often before a child enters grade school. These disorders are characterized by developmental deficits that impair personal, social, academic, or occupational functioning. Neurodevelopmental disorders include autism spectrum disorders (ASD), intellectual disabilities (ID), communication disorders, and attention-deficit/hyperactivity disorders (ADHD). These disorders, typically identified in early childhood, have common areas of involvement (comorbidity) and can significantly impact a child's development and learning ability. They vary in severity and may persist into adulthood, requiring ongoing support and intervention\cite{american2013diagnostic}. 

Attention deficits are often common in individuals with NDD due to the sensory system dysfunction that characterizes these disorders \cite{zomeren1994clinical}. Consequently, limited attention capability affects learning, social interactions, daily functioning, emotional regulation, safety, cognitive abilities, overall quality of life, and the ability to transfer knowledge from one circumstance to another. Strengthening attention skills can improve relationships, enhance independence, safer behavior, and a higher sense of well-being. This is why attention skills are often targeted in therapeutic interventions to support individuals in achieving their treatment goals.

Virtual Reality (VR) is achieving promising results in improving various aspects of neurodevelopmental disorders, including cognitive, motor, and social skills \cite{10.3389/fpsyt.2022.1055204,vona2020social,10.1007/978-3-030-50439-7_17} thanks to the possibility of creating a safe learning space where the user can train without unexpected events of the real environments while offering more engaging activities compared to traditional therapeutic ones. 

Virtual reality has also proven effective in the case of attention skills \cite{8944467,children9020250}, and in previous research, the potential of virtual reality to improve the attention skills of individuals with neurodevelopmental disorders was already been explored by designing Wildcard, an application for Google Cardboard\cite{10.1145/3078072.3084312}. With the arrival of more affordable virtual reality headsets and considering the various limitations of Wildcard related to the interaction paradigm (gaze-based, without controllers) and the graphic quality, it has been decided to upgrade Wildcard by moving from Google Cardboard to the Pico Neo3 Pro Eye, a standalone headset equipped with an eye-tracker and better performance and to run an exploratory study with the new version.
The objectives of the study were twofold:
\begin{enumerate}
    \item explore the usability of the new version of Wildcard with persons with NDD. Since the new version of Wildcard has a different interaction paradigm, it was necessary to verify that it was usable.
    \item explore the effectiveness of the new version of Wildcard in training attention skills. Since the primary goal of Wildcard is to improve attention skills, it was necessary to verify that the new interaction paradigm would lead to improvements in attention skills.
\end{enumerate}

\section{Related work}
\label{sec: works}
In the last decades, the literature has increasingly recognized the potential benefits of immersive technologies such as virtual reality in supporting the learning process and rehabilitation in individuals with NDD \cite{8010470,Tan2022} thanks to the possibility of creating a safe learning environment where the user can train without unexpected events of real environments. 
However, it is not easy to find applications that use virtual reality exclusively for attention skill enhancement; instead, these applications usually bring together multiple benefits that can improve the quality of life of people with NDD.

An example is the work of \cite{8944467} that shows how training joint attention skills can also benefit social and communication skills. This study is significant as joint attention, the ability to coordinate attention with another individual towards a common point of interest is challenging for many persons with ASD. The VR system simulates a classroom environment where children interact with avatars (virtual characters) that provide cues (like gaze direction, head-turning, and finger-pointing) directing attention to various objects. The children's task is to identify these objects based on the avatars’ cues correctly \cite{8944467}. 
Another example of how attention skills are trained with other abilities is the study of \cite{article}. It emphasizes recognizing affective states (like engagement, enjoyment, frustration, and boredom) through physiological signals during driving in VR \cite{article}. Attention skills are addressed indirectly through this framework. The VR environment requires participants to navigate and respond to various driving scenarios, implicitly engaging their attention skills. Although the primary focus isn't on attention skills per se, their treatment and enhancement are integral to the driving task and the study's broader goals \cite{article}.

In \cite{children9020250}, the authors utilize a VR classroom to create an immersive, controlled environment where children's responses to various stimuli can be observed and measured. It looks at gaze patterns as an indicator of attention skills, analyzing where and how long children focus their gaze during different classroom scenarios. This is crucial as gaze patterns in children with NDDs can significantly differ from typically developing children, impacting their learning and social interactions \cite{children9020250}.
Additionally, the study explores interoception – the awareness of internal body states – which is another key aspect of how children with NDDs process sensory information and react to their environment \cite{children9020250}. Another study involving a classroom scenario is the one by \cite{8780300} that presents an ADHD assessment using a virtual reality system. The core of the system is a simulated classroom environment designed to evaluate attention skills in individuals. The system's design integrates various elements to mimic real-life classroom distractions, allowing for a more accurate assessment of attention-related issues in ADHD. The methodology involves tracking behavioral responses and attention patterns in the VR setting, such as response time, accuracy, and focus consistency \cite{8780300}.

The integration of virtual reality with eye-tracking technology is giving good results and allows researchers and therapists to assess the condition of the user \cite{Selaskowski2023,Lee2023,Stokes2022,deBelen2023}, providing training tools to individuals with NDD \cite{Lee2020,McParland2021} for better practicing memory and attention skills.
A system that exploits eye tracking is also presented in \cite{9524690}, which is used for enhancing joint attention skills in children with Autism Spectrum Disorder (ASD). The system aims at fostering gaze sharing and gaze following, two critical components of joint attention \cite{9524690}. It encourages children to share their gaze by looking at an avatar's eye region and following the avatar's gaze to various objects in the virtual environment. This is significant because gaze sharing and following are essential for effective social communication but are often challenging for children with ASD \cite{9524690}. The system utilizes eye-tracking technology to monitor and analyze the child's gaze behavior, which is critical for assessing improvements in joint attention skills \cite{9524690}.

\section{Wildcard}
\begin{figure}[ht]
    \centering
    \includegraphics[width=1\textwidth]{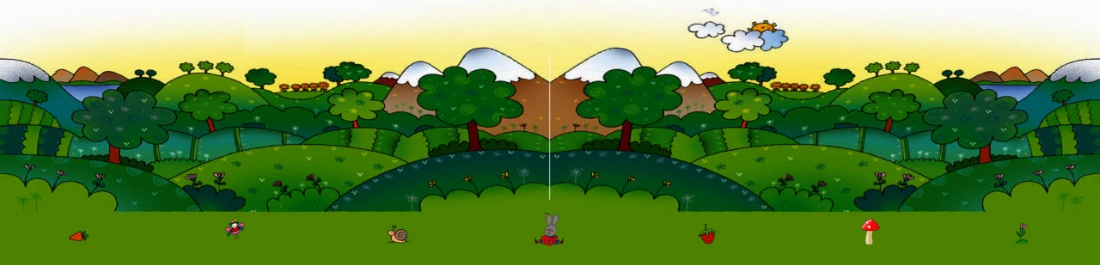}
    \caption{Example of a 360° environment for the \textit{Story} activity}
    \label{fig: story-example}
\end{figure}

In its first version, Wildcard was an immersive virtual reality mobile application running on smartphones and usable on Google Cardboard \cite{Garzotto2016}. Wildcard aims to improve people's attention skills with NDD through different gamified and customizable experiences. The activities contained in Wildcard were the output of an extensive collaboration with six caregivers, a game-pedagogy professor, and two computer scientists. For this reason, Wildcard tasks and graphics have been explicitly designed to promote attention. For example, i) requiring the child to point to an assigned visual element placed among multiple irrelevant ones, and ii) stabilizing their attention on an object and progressively changing direction to move it.
Caregivers suggested creating experiences that evoked the visual content of storybooks used in regular interventions . Familiar content was used to reduce the potential distress caused by unpredictable content. The stories have simple plots, a clear distinction between the main character and the subsidiary elements, and appropriate graphics (e.g., simple shapes, clear lines, and colors) (Figure \ref{fig: story-example}).

\subsection{Wildcard Activities}

The Wildcard experience encompasses three distinct activities: Research, Story, and Exploration. While the core design of these activities has been retained from the previous version, they have been improved to align with the new device's capabilities.

\begin{figure}[ht]
    \centering
    \begin{subfigure}[htbp]{0.45\textwidth}
         \centering
         \includegraphics[width=\textwidth]{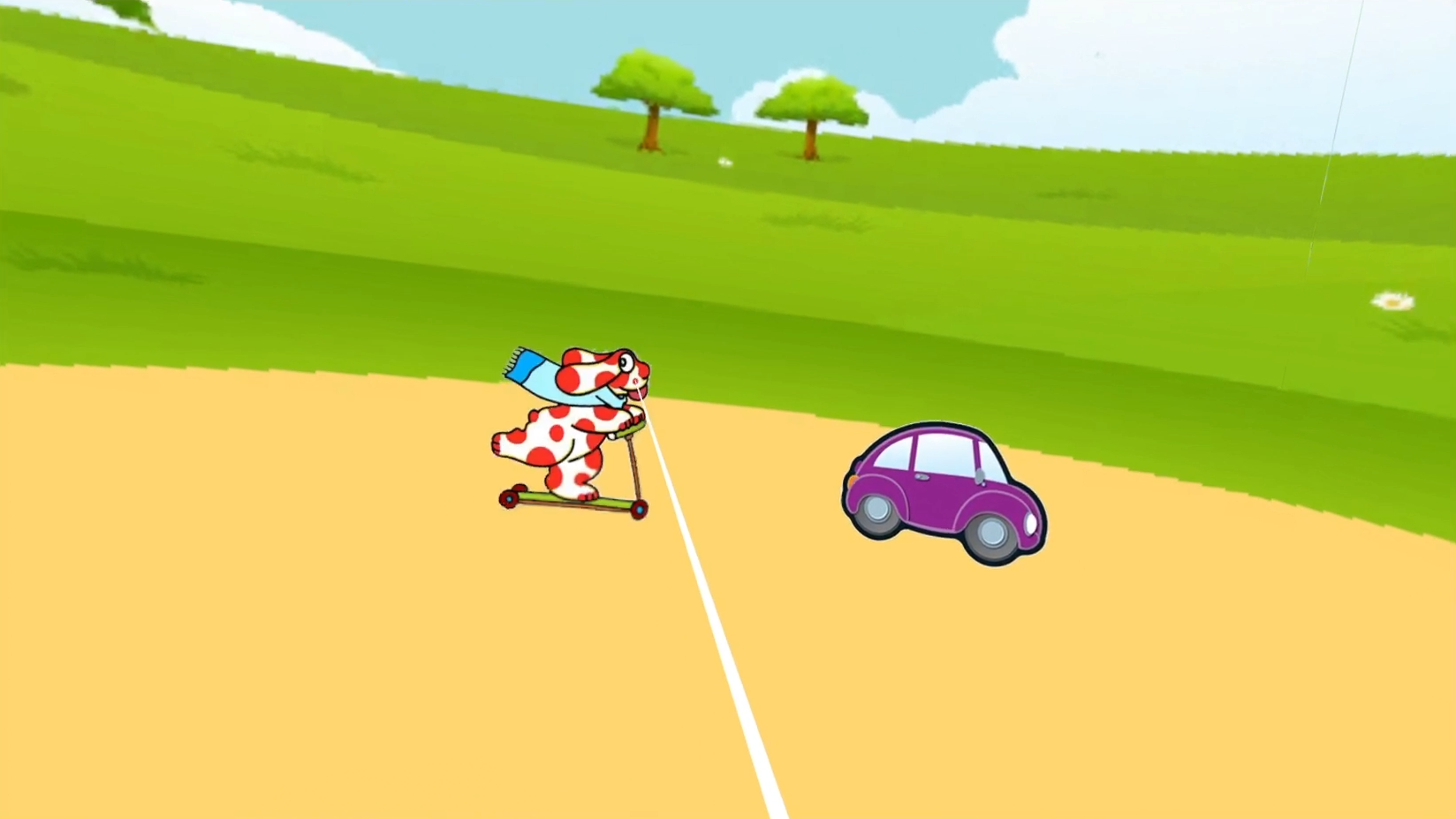}
        \caption{Story activity with la Pimpa}
        \label{fig: story}
     \end{subfigure}
     \hfill
     \begin{subfigure}[htbp]{0.5\textwidth}
         \centering
         \includegraphics[width=\textwidth]{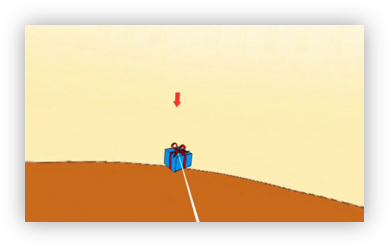}
         \caption{Object to find in Research activity}
         \label{fig: research}
     \end{subfigure}
    \caption{Examples from the Story and Research activities}
    \label{fig: researchstory}
\end{figure}

\paragraph{Research} In Research, users are tasked with locating and focusing on objects placed in the 360° environment, following the sequence of a story. The application guides users by highlighting the next target with an arrow to facilitate this task (Figure \ref{fig: research}). Once the correct object is identified, users employ a combination of gaze, controller pointing, and button press to interact, resulting in the disappearance of the targeted element and progression to the subsequent narrative element.

\paragraph{Story} The Story activity involves steering the experience's protagonist through the immersive environment, directing his actions while maintaining focus on him through gaze and controller pointing. The user's gaze initiates the constant movement of the protagonist, ensuring a measured pace that aligns with the narrative. This design choice reduces the need for users to accurately predict the protagonist's movements, promoting a more immersive and user-friendly experience (Figure \ref{fig: story}).

\begin{figure}[ht]
    \centering
    \begin{subfigure}[htbp]{0.65\textwidth}
         \centering
         \includegraphics[width=\textwidth]{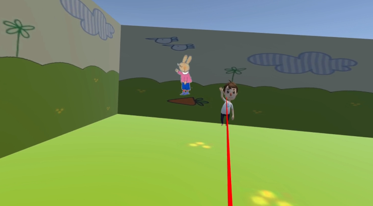}
         \caption{Example of a picture to point and the child support in the Exploration Activity}
     \end{subfigure}
     \hfill
     \begin{subfigure}[htbp]{0.3\textwidth}
         \centering
         \includegraphics[width=\textwidth]{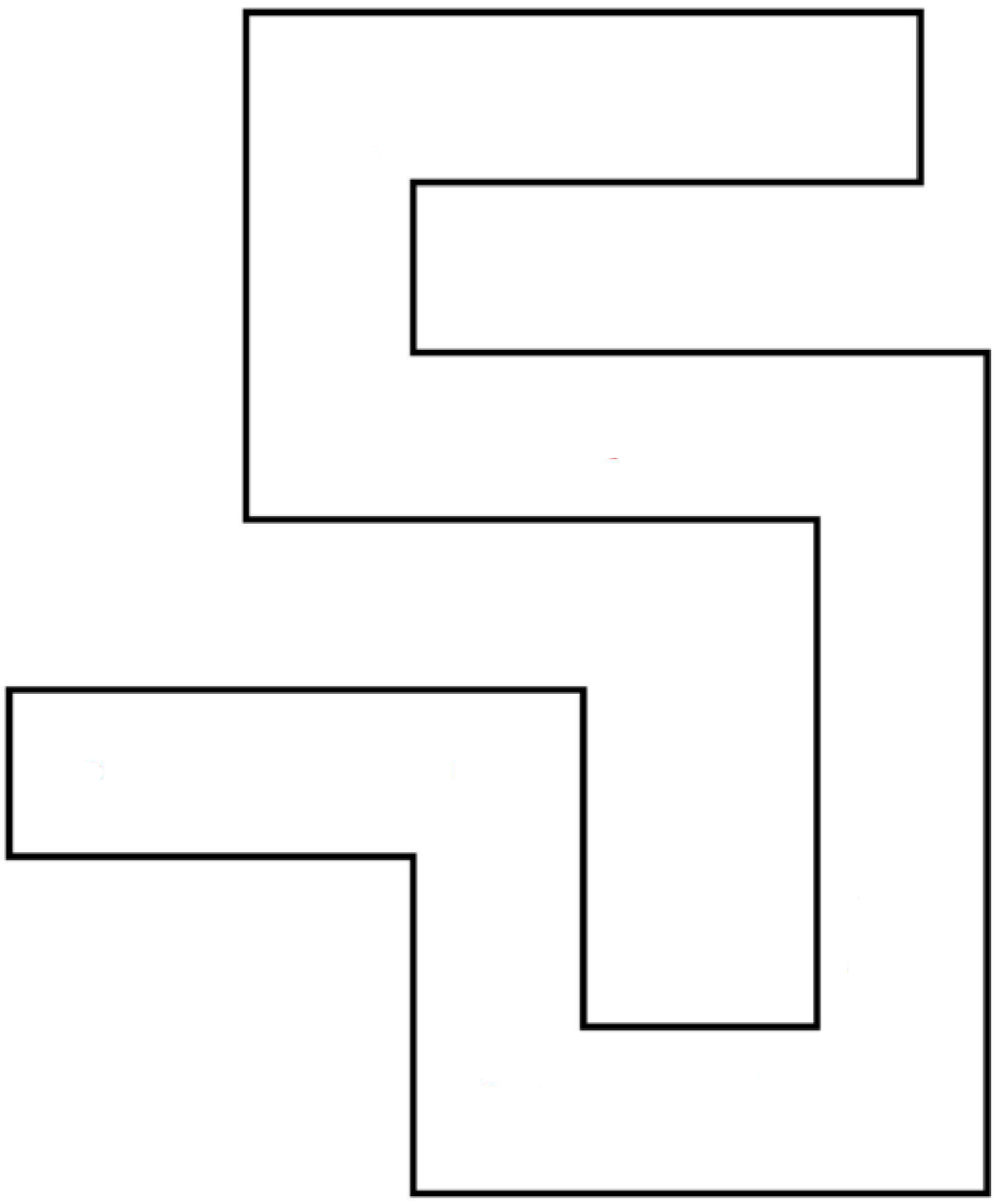}
         \caption{Map of the maze in the Exploration activity}
     \end{subfigure}
    \caption{Examples from the Exploration activity}
    \label{fig: maze}
\end{figure}

\paragraph{Exploration} In Exploration, users navigate a maze filled with story-related images on the walls. A model of a child is included as a companion to aid users in identifying the next picture of the story. Users can traverse the maze at a controlled, continuous pace using gaze and controller pointing to the correct picture, providing a well-integrated combination of exploration and narrative engagement (Figure \ref{fig: maze}).

\subsection{Interaction in Wildcard}
Previous user interaction within Wildcard was based on head orientation and focus. By changing head direction, users update their view of the virtual environment and have the illusion of moving in different directions. Since smartphone sensors (accelerometer and gyroscope) track only movement and orientation of the user’s head, the center of the smartphone screen was taken as a reference for the user's head (assumption: eye-gaze and head-gaze match). The center of the device was marked with a yellow dot (viewfinder). Users could control the viewfinder and interact with the system by moving their head. 
Although this system was usable by people with NDD the approximation that "head-gaze coincides with eye-gaze" was too strong for an application that is used to train attentional skills.
This is why, in its second iteration, Wildcard was designed for the Pico Neo3 Pro Eye, a VR HMD equipped with motion controllers and eye-tracking technology. This technological upgrade expanded the application's functionalities and introduced a more refined and immersive user experience. The integration of eye-tracking technology has proven critical, allowing for more detailed interactions within the virtual environment by precisely identifying the user's focal points. This can enhance the user's concentration on specific objectives, a crucial aspect of the proposed activities in Wildcard. To interact with the system, the user will not only have to look at the object they want to interact with but also point at it with the raycast of the controller. This additional action was added also to improve hand-eye coordination in these individuals which can equally contribute to improving attentional skills.

\section{Empirical Evaluation}
\subsection{Research Questions and Variables}
The main goals of our study were the evaluation of Wildcard usability and efficacy in enhancing user focus and attention, considering the introduction of different interaction methods in the new version. The research questions guiding this investigation are delineated as follows:
\begin{itemize}
    \item Is the Wildcard application usable?
    \item Is the Wildcard application effective in improving the attention skills of individuals with NDD?
\end{itemize}
By addressing these questions, the study aims to provide insights into the application's usability and to discern the efficacy of immersive experiences and eye-tracking technology in contributing to improve attentional skills among users with NDD.

From the research questions, three research variables with corresponding measurement tools were derived to obtain practical quantitative measures and responses, as shown in Table \ref{table: variables}.

\begin{table}
\centering
\begin{tabular}{|l|l|l|} 
\hline
\textbf{Research Question}                                                                                                                 & \textbf{Research Variable} & \textbf{Measurement tool}  \\ 
\hline
Is the Wildcard application usable?                                                                                                        & Usability                  & SUS                        \\ 
\hline
\multirow{2}{*}{\begin{tabular}[c]{@{}l@{}}Is the Wildcard application effective for\\enhancing attention in those with NDD?\end{tabular}} & Performance                & Completion Time                       \\ 
\cline{2-3}
                                                                                                                                           & Visual attention           & TMT                        \\
\hline
\end{tabular}
\caption{Research questions and associated research variables and measurement tools}
\label{table: variables}
\end{table}

\subsection{Participants}
A total of 38 adult users (age $\mu$: 30.89, $\sigma$: 8.13) were recruited for the study from a care center with a gender distribution of 32\% females and 68\% males. All participants have been diagnosed with a disorder falling within the spectrum of neurodevelopmental disorders, with their primary language proficiency being Italian. Additionally, to be eligible for the study, individuals had to be older than 16, exhibit cooperative behavior, demonstrate adequate verbal and vocal production capabilities, and concurrently demonstrate poor attentional stability. These criteria were established to ensure that participants possessed a baseline level of cognitive and communicative abilities while still representing the challenges associated with attentional deficits characteristic of NDD.

Conversely, exclusion criteria were implemented to maintain the integrity of the study sample. Individuals who exhibited oppositional behavior towards technology, aggression, hypersensitivity to visual and auditory stimuli, or poor hand-eye coordination were intentionally omitted from participation. These exclusion criteria were designed to mitigate factors that could impact the study's objectives, ensuring a more homogenous participant group and enhancing the reliability of the findings.

\subsection{Procedure}
The study unfolded over four weeks, structured within a three-step training protocol encompassing a pre-test, four weekly training sessions, and a concluding post-test. The sequential steps of the study were organized to assess the impact of the Wildcard application on individuals with NDD.

The beginning of the study involved a comprehensive pre-test phase where participants underwent an initial evaluation session under the supervision of caregivers. This session aimed to measure the baseline level of visual attention, employing the Trail Making Test (TMT) \cite{Bowie2006}. The TMT is a neuropsychological test focusing on visual attention and task-switching, requiring participants to connect 25 consecutive targets as rapidly as possible. It comprises two parts: connecting numbers 1 to 25 and alternating between letters and numbers (1-A-2-B...).

Following the pre-test, participants engaged in a structured training involving four sessions conducted at a frequency of one session per week. Researchers were present as observers, providing assistance and technical support during these sessions. Participants, seated on swivel chairs, were equipped with the Pico Neo3 Pro Eye device after the streaming of the user's view was enabled. This facilitated a real-time understanding of the participant's virtual environment, enabling caregivers and researchers to intervene if necessary.
Each session commenced with a calibration phase, utilizing the standard app provided by Pico to fine-tune the eye tracker according to the user's eyes. Subsequently, participants engaged in the Wildcard activities, following the sequence of \textit{Research}, \textit{Story}, and \textit{Exploration}. An essential aspect of the study design involved providing different stories for the activities in each session to prevent monotony and repetition, thereby enhancing participant engagement.

At the end of the first session, participants were invited to respond to the SUS questionnaire to gather valuable data and feedback concerning the system's usability and other relevant information. The four consecutive sessions were concluded by a post-test, mirroring the TMT conducted during the pre-test to gather more information about possible improvements.

\subsection{Data gathering methods}
The study employed multiple methods to assess the impact and efficacy of the Wildcard application. The data-gathering methods are delineated as follows:

\begin{itemize}
    \item \textbf{Usability}: the System Usability Scale (SUS) \cite{SUS} was utilized to gather data concerning the usability of the system. This standardized questionnaire comprises ten Likert-scale questions designed to quantitatively measure the users' perceptions and satisfaction levels regarding the Wildcard application's usability. The final scores were computed using the formula:
    $$(\sum_{n=1}^{5} (Score_{2n-1} - 1) + \sum_{n=1}^{5} (5 - Score_{2n})) * 2.5$$

    \item \textbf{Performance}: the time required for participants to complete each activity during the training sessions was recorded to gather insights into the participants' engagement and proficiency with the Wildcard application. 
    
    \item \textbf{Visual attention}: the Trail Making Test was a fundamental tool for evaluating the level of visual attention. By comparing the pre-test and post-test scores, it is possible to quantify any improvements resulting from the Wildcard training sessions.
\end{itemize}

\section{Results and Discussion}
\subsection{Usability}
Of the 38 users involved in the study, only 37 answers have been considered for the evaluation of the usability of Wildcard through the SUS questionnaire. This exclusion was necessary because one participant's responses were considered nonsensical since he consistently provided the lowest score for all questions, irrespective of their nature, making the evaluation unreliable and inconsistent.

\begin{figure}
    \centering
    \includegraphics[width=1\textwidth]{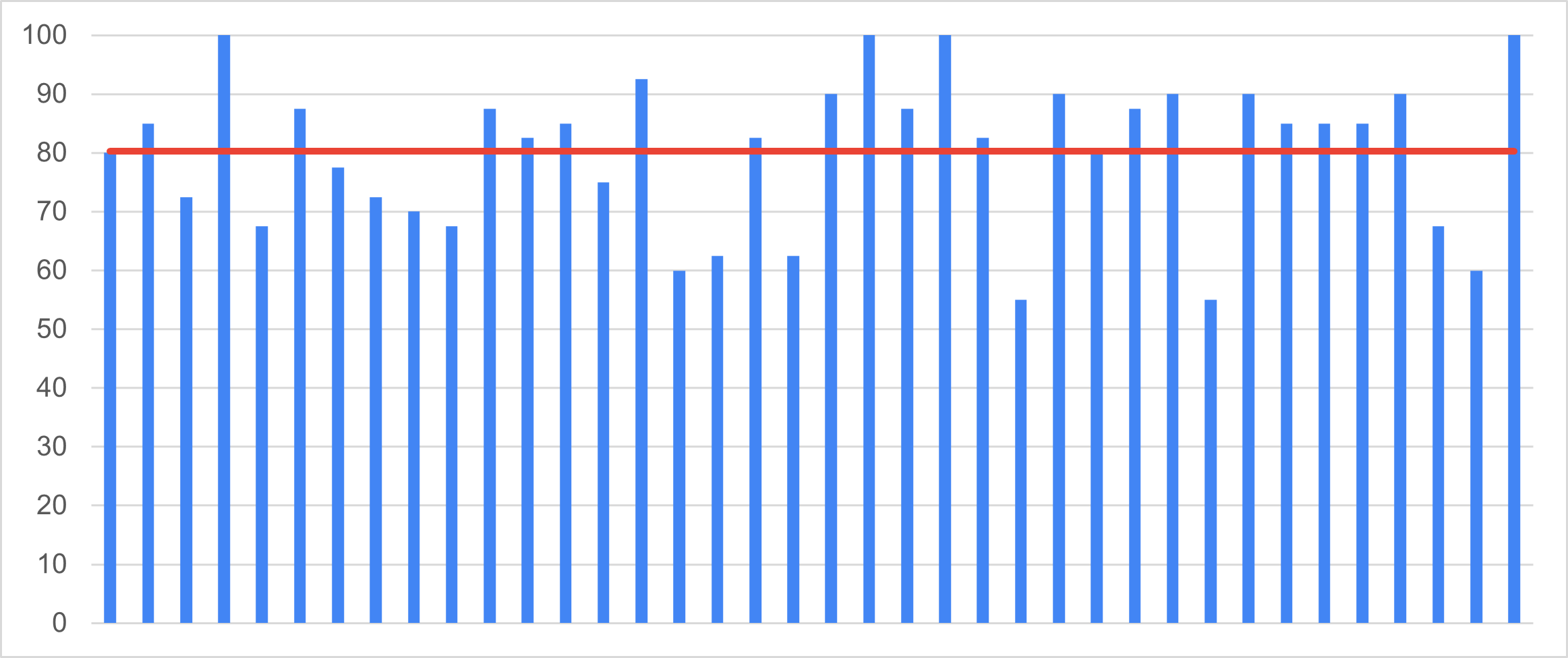}
    \caption{SUS: user and average scores}
    \label{fig:SUS}
\end{figure}

Despite this isolated instance, the overall analysis of the collected usability data shows a positive picture regarding the Wildcard application's usability for individuals with NDD. The average SUS score, calculated from the valid responses, is an encouraging 80.26 (Figure \ref{fig:SUS}). According to the interpretative guidelines by Bangor et al. \cite{SUSEval}, this score falls within the "Good" interval. This denotes that, on average, participants perceived the Wildcard application as user-friendly and navigable, indicating a positive usability experience within the group of individuals with NDD.

\subsection{Performance}
\begin{figure}[htbp]
    \centering
    \begin{subfigure}[htbp]{0.326\textwidth}
        \includegraphics[width=\textwidth]{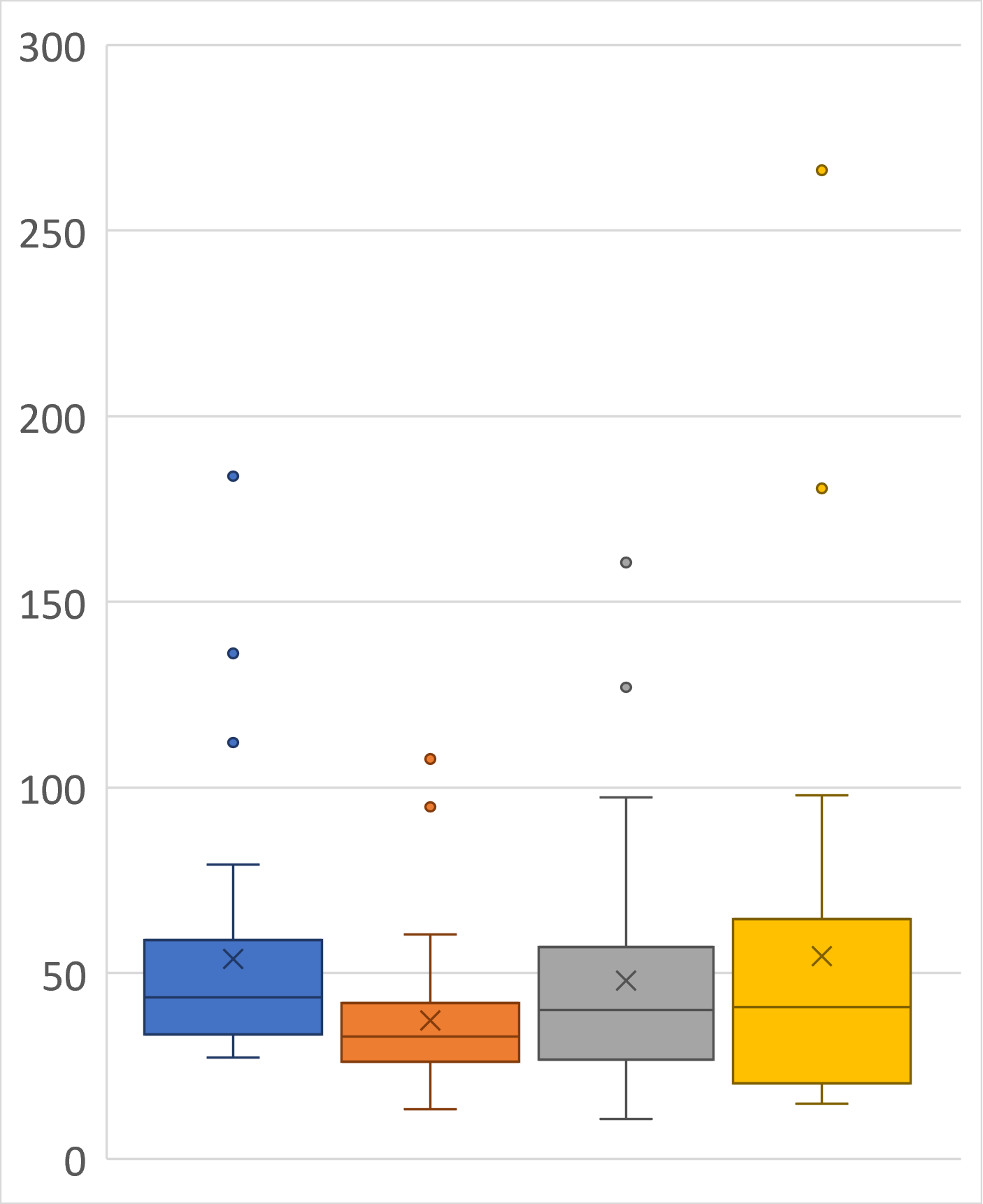}
        \caption{Research}
    \end{subfigure}
    \begin{subfigure}[htbp]{0.326\textwidth}
        \includegraphics[width=\textwidth]{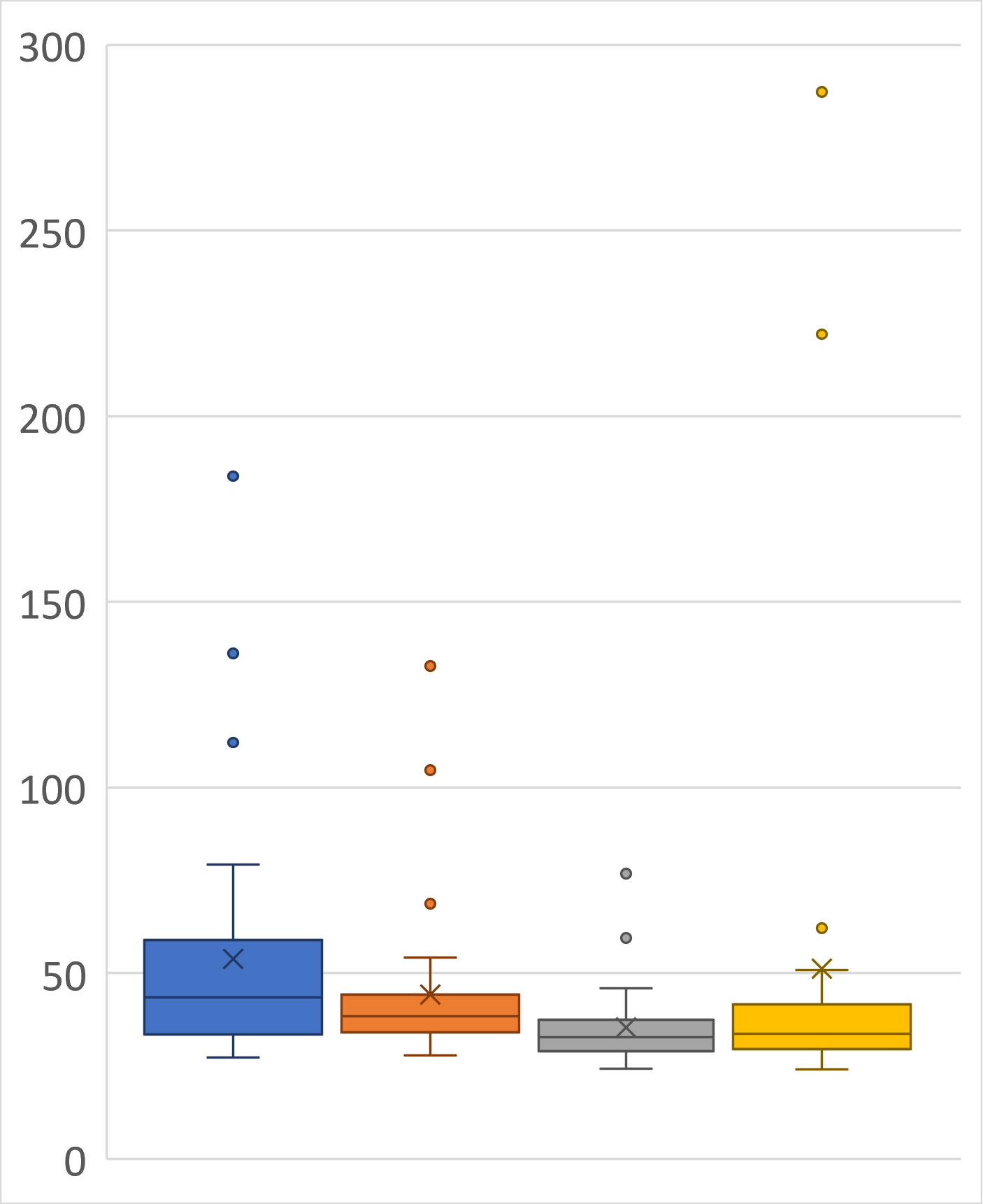}
        \caption{Story}
    \end{subfigure}
    \begin{subfigure}[htbp]{0.326\textwidth}
        \includegraphics[width=1\textwidth]{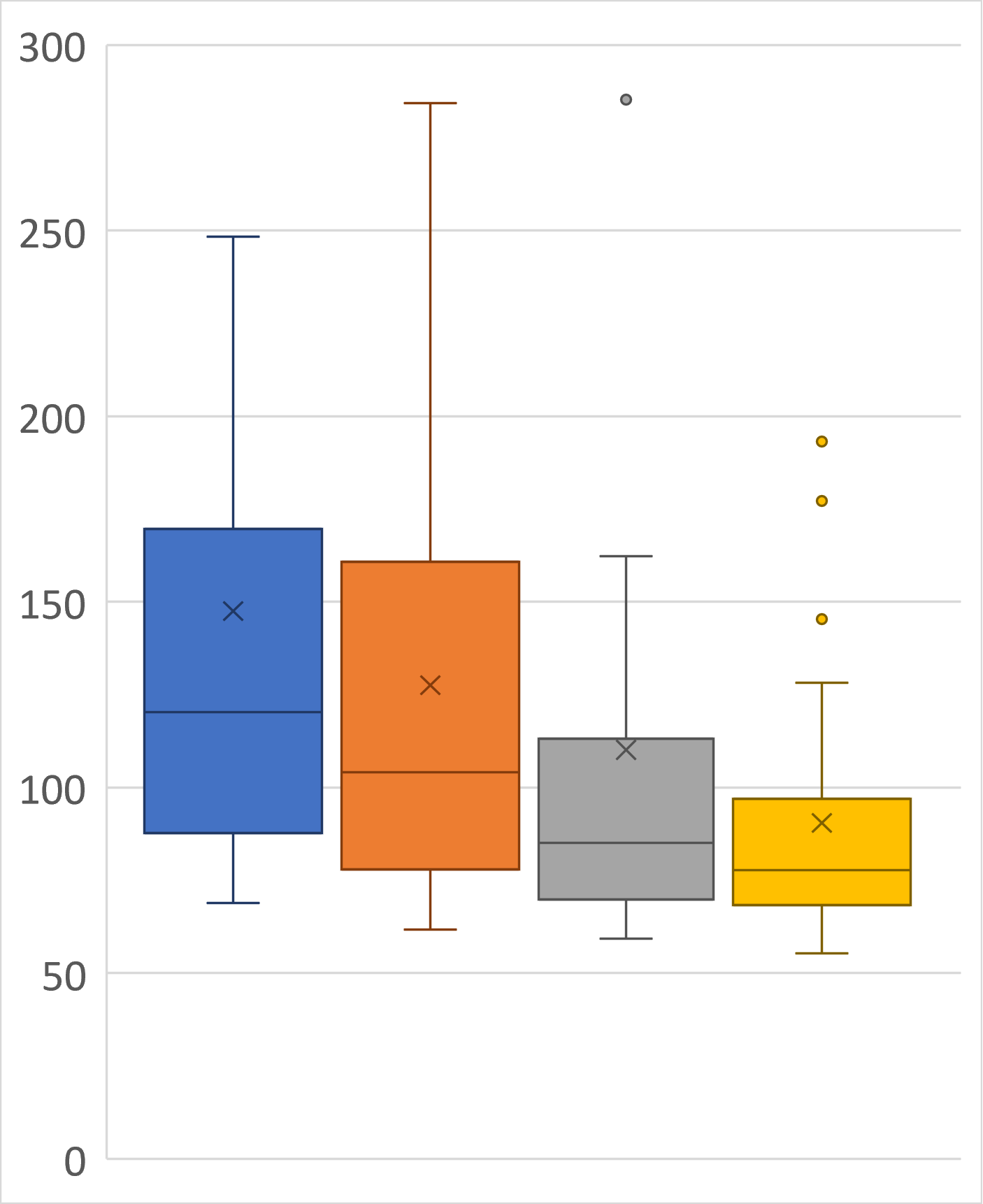}
        \caption{Exploration}
    \end{subfigure}
    \caption{Time required to execute the activities (in seconds)}
    \label{fig: performance}
\end{figure}

Figure \ref{fig: performance} illustrates the trend in the overall time required to complete each task across the sessions, showing a consistent decrease in completion times after each session. This observed pattern signifies a tangible improvement in task completion efficiency as participants engaged with the Wildcard application over successive sessions.

Notably, the \textit{Research} activity slightly differed from this trend. However, this difference can be attributed to a deliberate increase in the number of elements within the environment after each session. As the complexity of the task increased with additional elements, the marginal decrease in completion times for this specific activity aligns with the logical expectation that more elements demand a proportionally longer time to complete the task.

Additionally, it is important to notice that, despite the overall positive trends in task completion, organizational challenges and technical issues with the eye tracker resulted in a decline in participant participation across successive sessions. While most of the 38 users successfully completed the first session, logistical constraints and unanticipated problems with the eye tracker hindered continued participation for some individuals in subsequent sittings. 

\subsection{Visual Attention}
A selective approach was employed in analyzing the study data, considering participants who consistently engaged with the Wildcard application. Specifically, only those who completed a minimum of three out of the four Wildcard test sessions were included, resulting in an evaluation group of 27 participants. This decision aimed to ensure that the analysis encompassed individuals with substantial and consistent experience with the application, enhancing the reliability and value of the findings.

\begin{figure}
    \centering
    \includegraphics[width=1\textwidth]{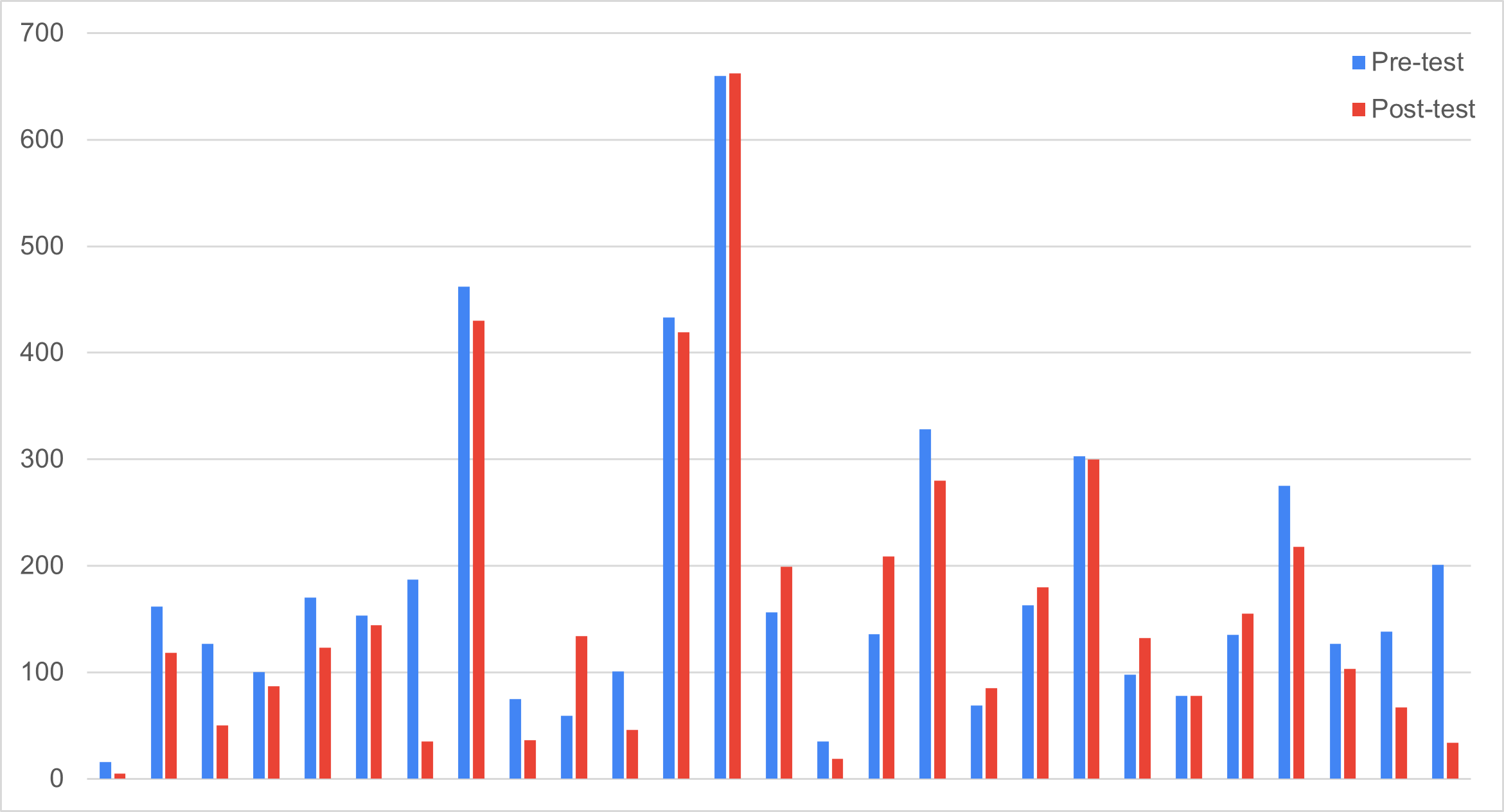}
    \caption{TMT: pre-test and post-test results}
    \label{fig:TMT}
\end{figure}

The analysis revealed a positive trend (Figure \ref{fig:TMT}), with most participants demonstrating improvement in the time required to complete the TMT. Out of the 27 participants, only six did not exhibit a reduction in completion time. This consistent improvement was also evident in the average completion time, which decreased from 183.22 seconds to 161.04 seconds.

\section{Conclusion}

The findings of the proposed study suggest that Wildcard and, more broadly, virtual reality exploiting eye-tracking technologies enable the development of personalized and targeted interventions tailored to the needs of individuals with neurodevelopmental disorders. The positive results observed in visual attention improvement highlight these technologies' potential to address attentional challenges associated with NDD, creating opportunities for more effective and customized therapeutic interventions.


These preliminary results do not prove that Wildcard is a therapeutic tool. However, this study serves as a stepping stone, encouraging further exploration and refinement of these innovative approaches to contribute to the well-being and cognitive development of individuals facing challenges in attentional stability.

\subsection{Limitations}
Despite the promising outcomes of the study highlighting the potential of Wildcard in enhancing visual attention among individuals with NDD, several limitations must be acknowledged.


Firstly, the heterogeneity within the NDD spectrum poses a challenge in understanding the effectiveness of Wildcard on a specific group of disorders. Therefore, it is necessary to involve more individuals with the same diagnosis to offer more specific results.

Secondly, the study encountered challenges related to organizational and technological issues, impacting the completion of sessions within the four weeks. Despite efforts to reschedule sessions, logistical constraints occasionally led to skipped meetings, potentially affecting the overall reliability of the findings.

A notable concern also arose from difficulties in assisting participants when interactions occurred through the eye tracker embedded in the HMD. While caregivers could address issues related to the controller by guiding users, assisting with the eye tracker proved more challenging. Even with screen casting to display the user's perspective, direct assistance remained impossible, leading to certain activities taking longer than expected.

\subsection{Future Works}
While acknowledging the limitations, the study's positive outcomes show that it is possible to improve attentional abilities in individuals with NDD by leveraging VR and eye-tracking technology. To further improve the validity of the study and the usage of these technologies, some considerations for future research can be considered.

Given the difficulties in providing direct assistance during eye-tracking interactions, possible technology support could be explored. Developing more user-friendly interfaces, providing a visible indicator of the user's focus (possibly visible only by the caregiver through the screen casting) or incorporating real-time assistance features could enhance the overall experience for individuals with NDD.

Moreover, further studies will be carried out with stricter inclusion criteria in order to recruit a more homogeneous sample. This approach would contribute to a deeper understanding of whether Wildcard can improve attention skills on specific kinds of NDD.

\begin{credits}
\subsubsection{\ackname}
This work is partially funded by TIM S.p.A. through its UniversiTIM granting program.
\subsubsection{\discintname}
The authors have no competing interests to declare that are relevant to the content of this article.
\end{credits}
%
%
%
\bibliographystyle{splncs04}
\bibliography{main.bib}

\begin{thebibliography}{10}
\providecommand{\url}[1]{\texttt{#1}}
\providecommand{\urlprefix}{URL }
\providecommand{\doi}[1]{https://doi.org/#1}

\bibitem{9524690}
Amat, A.Z., Zhao, H., Swanson, A., Weitlauf, A.S., Warren, Z., Sarkar, N.: Design of an interactive virtual reality system, invirs, for joint attention practice in autistic children. IEEE Transactions on Neural Systems and Rehabilitation Engineering  \textbf{29},  1866--1876 (2021). \doi{10.1109/TNSRE.2021.3108351}

\bibitem{american2013diagnostic}
American Psychiatric~Association, D., Association, A.P., et~al.: Diagnostic and statistical manual of mental disorders: DSM-5, vol.~5. American psychiatric association Washington, DC (2013)

\bibitem{SUSEval}
Bangor, A., Kortum, P., Miller, J.: Determining what individual sus scores mean: adding an adjective rating scale. J. Usability Studies  \textbf{4}(3),  114–123 (may 2009)

\bibitem{deBelen2023}
de~Belen, R.A., Pincham, H., Hodge, A., Silove, N., Sowmya, A., Bednarz, T., Eapen, V.: Eye-tracking correlates of response to joint attention in preschool children with autism spectrum disorder. {BMC} Psychiatry  \textbf{23}(1) (Mar 2023). \doi{10.1186/s12888-023-04585-3}, \url{https://doi.org/10.1186/s12888-023-04585-3}

\bibitem{article}
Bian, D., Wade, J., Swanson, A., Warren, Z., Sarkar, N.: Physiology-based affect recognition during driving in virtual environment for autism intervention. PhyCS 2015 - 2nd International Conference on Physiological Computing Systems, Proceedings pp. 137--145 (01 2015)

\bibitem{Bowie2006}
Bowie, C.R., Harvey, P.D.: Administration and interpretation of the trail making test. Nature Protocols  \textbf{1}(5),  2277–2281 (Dec 2006). \doi{10.1038/nprot.2006.390}, \url{http://dx.doi.org/10.1038/nprot.2006.390}

\bibitem{8010470}
Bozgeyikli, L., Raij, A., Katkoori, S., Alqasemi, R.: A survey on virtual reality for individuals with autism spectrum disorder: Design considerations. IEEE Transactions on Learning Technologies  \textbf{11}(2),  133--151 (2018). \doi{10.1109/TLT.2017.2739747}

\bibitem{SUS}
Brooke, J.: Sus: A quick and dirty usability scale. Usability Eval. Ind.  \textbf{189} (11 1995)

\bibitem{Garzotto2016}
Garzotto, F., Gelsomini, M., Clasadonte, F., Montesano, D., Occhiuto, D.: Wearable immersive storytelling for disabled children. In: Proceedings of the International Working Conference on Advanced Visual Interfaces. {ACM} (Jun 2016). \doi{10.1145/2909132.2909256}, \url{https://doi.org/10.1145/2909132.2909256}

\bibitem{10.1145/3078072.3084312}
Garzotto, F., Gelsomini, M., Occhiuto, D., Matarazzo, V., Messina, N.: Wearable immersive virtual reality for children with disability: a case study. In: Proceedings of the 2017 Conference on Interaction Design and Children. p. 478–483. IDC '17, Association for Computing Machinery, New York, NY, USA (2017). \doi{10.1145/3078072.3084312}, \url{https://doi.org/10.1145/3078072.3084312}

\bibitem{children9020250}
Ide-Okochi, A., Matsunaga, N., Sato, H.: A preliminary study of assessing gaze, interoception and school performance among children with neurodevelopmental disorders: The feasibility of vr classroom. Children  \textbf{9}(2) (2022). \doi{10.3390/children9020250}, \url{https://www.mdpi.com/2227-9067/9/2/250}

\bibitem{8944467}
Jyoti, V., Gupta, S., Lahiri, U.: Virtual reality based avatar-mediated joint attention task for children with autism: Implication on performance and physiology. In: 2019 10th International Conference on Computing, Communication and Networking Technologies (ICCCNT). pp.~1--7 (2019). \doi{10.1109/ICCCNT45670.2019.8944467}

\bibitem{Lee2023}
Lee, D.Y., Shin, Y., Park, R.W., Cho, S.M., Han, S., Yoon, C., Choo, J., Shim, J.M., Kim, K., Jeon, S.W., Kim, S.J.: Use of eye tracking to improve the identification of attention-deficit/hyperactivity disorder in children. Scientific Reports  \textbf{13}(1) (Sep 2023). \doi{10.1038/s41598-023-41654-9}, \url{https://doi.org/10.1038/s41598-023-41654-9}

\bibitem{Lee2020}
Lee, T., Yeung, M., Sze, S., Chan, A.: Computerized eye-tracking training improves the saccadic eye movements of children with attention-deficit/hyperactivity disorder. Brain Sciences  \textbf{10}(12), ~1016 (Dec 2020). \doi{10.3390/brainsci10121016}, \url{https://doi.org/10.3390/brainsci10121016}

\bibitem{McParland2021}
McParland, A., Gallagher, S., Keenan, M.: Investigating gaze behaviour of children diagnosed with autism spectrum disorders in a classroom setting. Journal of Autism and Developmental Disorders  \textbf{51}(12),  4663--4678 (Feb 2021). \doi{10.1007/s10803-021-04906-z}, \url{https://doi.org/10.1007/s10803-021-04906-z}

\bibitem{10.1007/978-3-030-50439-7_17}
Nordahl-Hansen, A., Dechsling, A., S{\"u}tterlin, S., B{\o}rtveit, L., Zhang, D., {\O}ien, R.A., Marschik, P.B.: An overview of virtual reality interventions for two neurodevelopmental disorders: Intellectual disabilities and autism. In: Schmorrow, D.D., Fidopiastis, C.M. (eds.) Augmented Cognition. Human Cognition and Behavior. pp. 257--267. Springer International Publishing, Cham (2020)

\bibitem{Selaskowski2023}
Selaskowski, B., Asch{\'{e}}, L.M., Wiebe, A., Kannen, K., Aslan, B., Gerding, T.M., Sanchez, D., Ettinger, U., K\"{o}lle, M., Lux, S., Philipsen, A., Braun, N.: Gaze-based attention refocusing training in virtual reality for adult attention-deficit/hyperactivity disorder. {BMC} Psychiatry  \textbf{23}(1) (Jan 2023). \doi{10.1186/s12888-023-04551-z}, \url{https://doi.org/10.1186/s12888-023-04551-z}

\bibitem{Stokes2022}
Stokes, J.D., Rizzo, A., Geng, J.J., Schweitzer, J.B.: Measuring attentional distraction in children with {ADHD} using virtual reality technology with eye-tracking. Frontiers in Virtual Reality  \textbf{3} (Mar 2022). \doi{10.3389/frvir.2022.855895}, \url{https://doi.org/10.3389/frvir.2022.855895}

\bibitem{10.3389/fpsyt.2022.1055204}
Tan, B.L., Shi, J., Yang, S., Loh, H., Ng, D., Choo, C., Medalia, A.: The use of virtual reality and augmented reality in psychosocial rehabilitation for adults with neurodevelopmental disorders: A systematic review. Frontiers in Psychiatry  \textbf{13} (2022). \doi{10.3389/fpsyt.2022.1055204}, \url{https://www.frontiersin.org/journals/psychiatry/articles/10.3389/fpsyt.2022.1055204}

\bibitem{Tan2022}
Tan, B.L., Shi, J., Yang, S., Loh, H., Ng, D., Choo, C., Medalia, A.: The use of virtual reality and augmented reality in psychosocial rehabilitation for adults with neurodevelopmental disorders: A systematic review. Frontiers in Psychiatry  \textbf{13} (Dec 2022). \doi{10.3389/fpsyt.2022.1055204}, \url{https://doi.org/10.3389/fpsyt.2022.1055204}

\bibitem{8780300}
Tan, Y., Zhu, D., Gao, H., Lin, T.W., Wu, H.K., Yeh, S.C., Hsu, T.Y.: Virtual classroom: An adhd assessment and diagnosis system based on virtual reality. In: 2019 IEEE International Conference on Industrial Cyber Physical Systems (ICPS). pp. 203--208 (2019). \doi{10.1109/ICPHYS.2019.8780300}

\bibitem{vona2020social}
Vona, F., Silleresi, S., Beccaluva, E., Garzotto, F.: Social matchup: Collaborative games in wearable virtual reality for persons with neurodevelopmental disorders. In: Joint International Conference on Serious Games. pp. 49--65. Springer (2020)

\bibitem{zomeren1994clinical}
Zomeren, A.H., Brouwer, W.H.: Clinical neuropsychology of attention. Oxford University Press, USA (1994)

\end{thebibliography}
\end{document}